\documentclass[%
 reprint,
superscriptaddress,
 amsmath,amssymb,
 aps, prl,
]{revtex4-1}

\usepackage{graphicx} 
\usepackage{dcolumn} 
\usepackage{sidecap}
\usepackage{bm} 
\usepackage[svgnames]{xcolor}
\usepackage[bookmarks=false,linkcolor=blue,urlcolor=blue,colorlinks,citecolor=blue]{hyperref}
\begin{document}

\title{Control of Andreev bound states using superconducting phase texture} 

\author{Abhishek Banerjee}
\affiliation{Center for Quantum Devices, Niels Bohr Institute,
University of Copenhagen,
Universitetsparken 5, 2100 Copenhagen, Denmark
}

\author{Max Geier}
\affiliation{Center for Quantum Devices, Niels Bohr Institute,
University of Copenhagen,
Universitetsparken 5, 2100 Copenhagen, Denmark
}

\author{Md Ahnaf Rahman}
\affiliation{Center for Quantum Devices, Niels Bohr Institute,
University of Copenhagen,
Universitetsparken 5, 2100 Copenhagen, Denmark
}

\author{Daniel S.~Sanchez}
\affiliation{Center for Quantum Devices, Niels Bohr Institute,
University of Copenhagen,
Universitetsparken 5, 2100 Copenhagen, Denmark
}

\author{Candice Thomas}
\affiliation{Department of Physics and Astronomy, and Birck Nanotechnology Center,
Purdue University, West Lafayette, Indiana 47907 USA
}

\author{Tian Wang}
\affiliation{Department of Physics and Astronomy, and Birck Nanotechnology Center,
Purdue University, West Lafayette, Indiana 47907 USA
}

\author{Michael J.~Manfra}
\affiliation{Department of Physics and Astronomy, and Birck Nanotechnology Center,
Purdue University, West Lafayette, Indiana 47907 USA
}

\affiliation{School of Materials Engineering, and Elmore Family School of Electrical and Computer Engineering, Purdue University, West Lafayette, Indiana 47907 USA}

\author{Karsten Flensberg}
\affiliation{Center for Quantum Devices, Niels Bohr Institute,
University of Copenhagen,
Universitetsparken 5, 2100 Copenhagen, Denmark
}

\author{Charles~M.~Marcus}
\affiliation{Center for Quantum Devices, Niels Bohr Institute,
University of Copenhagen,
Universitetsparken 5, 2100 Copenhagen, Denmark
} 

\date{\today} 

\begin{abstract}
Andreev bound states with opposite phase-inversion asymmetries are observed in local and nonlocal tunneling spectra at the two ends of a superconductor-semiconductor-superconductor planar Josephson junction in the presence of a perpendicular magnetic field. Spectral signatures agree with a theoretical model, yielding a physical picture in which phase textures in superconducting leads localize and control the position of Andreev bound states in the junction, demonstrating a simple means of controlling the position and size of Andreev states within a planar junction.
\end{abstract}

\maketitle
The manipulation of spatial properties of Andreev bound states (ABSs), both their position and spatial extent, is an important goal in contemporary superconducting physics. Braiding protocols that can test exotic exchange statistics, including constructions of a topological quantum processor, rely on spatially exchanging the positions of special types of ABSs, such as Majorana zero modes (MZMs)~\cite{Nayak2008,alicea2011,Stern2013,Kitaev2003}. The spatial extent of ABSs is a key parameter that determines the coupling of states in Andreev molecules~\cite{pillet2019,Kornich2019}, Majorana chains, and quantum-dot states interacting via superconductors~\cite{choy2011majorana,sau2012realizing,Pientka2013topological}. A general method for spatially manipulating ABSs, that is also fast and hysteresis-free, is highly sought-after. In this direction, previous works have suggested the use of chemical potential~\cite{alicea2011} and magnetic field texture~\cite{KjaergaardPRB2012, KlinovajaPRX2013, FatinWireless2016} for spatial control of ABSs. Experimentally, an array of electrostatic gates has been used for local ABS control~\cite{Elfeky2021,poschl2022nonlocal}. 

Controllable superconducting phase provides an additional useful knob for ABS manipulation. In bulk superconductors, Abrikosov and Pearl vortices~\cite{abrikosov1957magnetic,pearl1964current} represent windings of the superconducting phase that can trap ABSs, including MZMs~\cite{Ivanov2001,FuKane2008}. Spatial manipulation of individual vortices has been demonstrated~\cite{gardner2002manipulation,straver2008controlled,auslaender2009mechanics,Shapira2015,Kalisky2016,ge2016nanoscale} . Similarly, Josephson vortices arising from phase windings in superconductor-normal-superconductor (SNS) junctions, are also amenable to spatial manipulation~\cite{tinkham_introduction_2004,likharev1979superconducting,roditchev2015direct,carapella2016current,dremov2019local} and can host MZMs~\cite{grosfeld2011observing,Stern2019Fractional,Ma2020Braiding}. In planar Josephson junctions (PJJs), recent proposals have suggested that superconducting phase textures, not necessarily in a vortex configuration, can be used to spatially control MZMs and even execute braiding operations~\cite{Stern2019Fractional,Ma2020Braiding,ShabaniXjunction}. Encouraging progress has been made towards realizing topological superconductivity on this platform~\cite{HellFlensberg,HellPRB2017,Pientka, fornieri,ren,ShabaniReopening,GoswamiReopening,GapReopening}. 

In this Letter, we study PJJs consisting of superconductor-normal-superconductor (SNS) junctions, where N is a semiconductor with strong spin-orbit coupling. We focus on non-topological ABSs and study their response to a perpendicular magnetic field applied through the junction. A spatially varying phase texture is induced on the two superconducting leads. ABSs formed in the N region respond to this phase texture such that their localization length is controlled by the magnitude of magnetic flux penetrating the junction area and their position is controlled by a phase bias applied across the junction. 

The PJJs are fabricated on InAs/Al heterostructure stacks. The N region comprises an InAs layer with Al stripped away, whereas the S regions are composed of patterned Al/InAs superconducting leads. Quantum point contacts, formed by electrostatic gating, allow tunneling spectroscopy at the top and bottom ends of the junction. Nonlocal electrical transport between the two ends allows bulk spectroscopy~\cite{Akhmerov,DanonNonlocal,MenardNonlocal,AnselmettiNonLocal,BanerjeeNonlocal}. Phase-biasing of the junction is obtained by embedding in a radio-frequency superconducting quantum interference device (rf-SQUID) geometry, consisting of a superconducting loop. At low perpendicular magnetic field, $|B_\perp| \sim$~0.1~mT, we observe phase-inversion-symmetric conductance spectra. As $|B_\perp|$ is increased, the local conductance spectra become phase-inversion-asymmetric, with opposite phase-asymmetries in the top and bottom local conductances. On the other hand, the nonlocal conductance spectrum remains relatively phase-symmetric within each flux lobe. 

\begin{figure*}[t]
\includegraphics[width=1\textwidth]{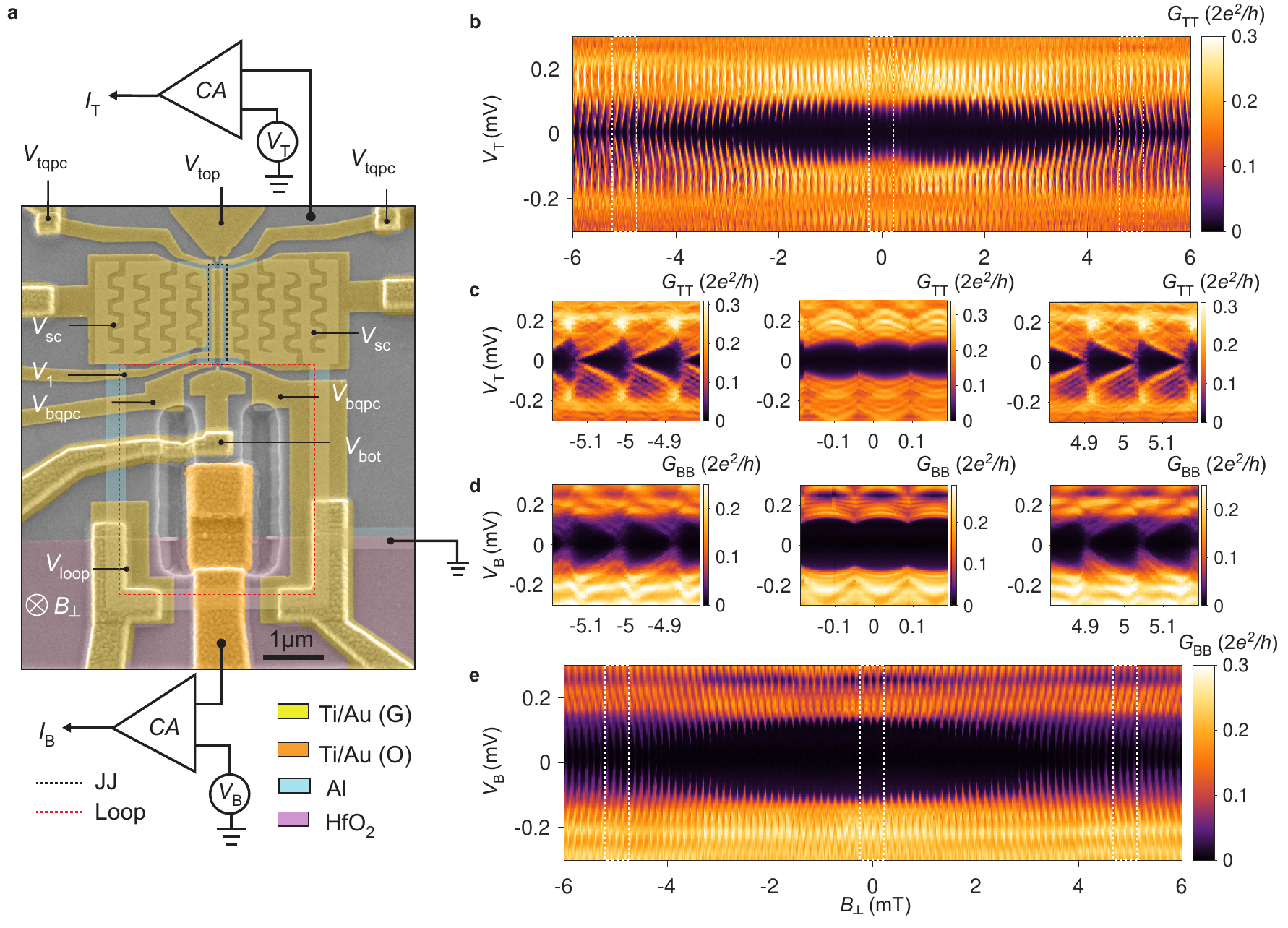}
\caption{\label{fig01} {\bf Device micrograph and differential conductance as a function of perpendicular magnetic field.}  (a)~False-color electron micrograph of a planar Josephson junction device measured in a three-terminal configuration allowing local and nonlocal tunneling spectroscopy. DC biases, $V_{\rm T}$ and $V_{\rm B}$, are applied to top and bottom ohmic contacts through current amplifiers (CA) connected to the respective terminals. The superconducting loop is grounded. Gates $V_{\rm top(bot)}$ and $V_{\rm t(b)qpc}$ create an electrostatic constriction at the top (bottom) end for tunneling spectroscopy. $V_{\rm sc}$ controls density under the superconducting leads. $V_{1}$ controls density in the junction. An out-of-plane magnetic field threads magnetic flux through the device. Two distinct areas of flux penetration are identified: the Josephson junction (JJ) indicated by a black dashed rectangle, and the superconducting loop indicated by red dashed rectangle of larger size. (b) and (e) Differential conductance measured at the top and bottom ends of the junction, as a function of the out-of-plane magnetic field $B_\perp$. The superconducting gap oscillates periodically, with period comparable to $\Phi_0 = h/2e$ through the superconducting loop. The magnitude of the gap at both ends is diminished as $|B_\perp|$ increases. (c) Top and (b) Bottom end differential conductance spectrum showing $\sim$~3 flux lobes centered around (left) $B_\perp=-5$~mT (center) $B_\perp=0$ and (right) $B_\perp=+5$~mT. Notice that the top and bottom conductance spectra become phase-asymmetric at finite $B_\perp$, with the sense of asymmetry reversed upon changing the sign of $B_\perp$. The sense of asymmetry is opposite for the top and bottom conductance spectra.}
\end{figure*}

\begin{figure*}[t]
\includegraphics[width=1.0\textwidth]{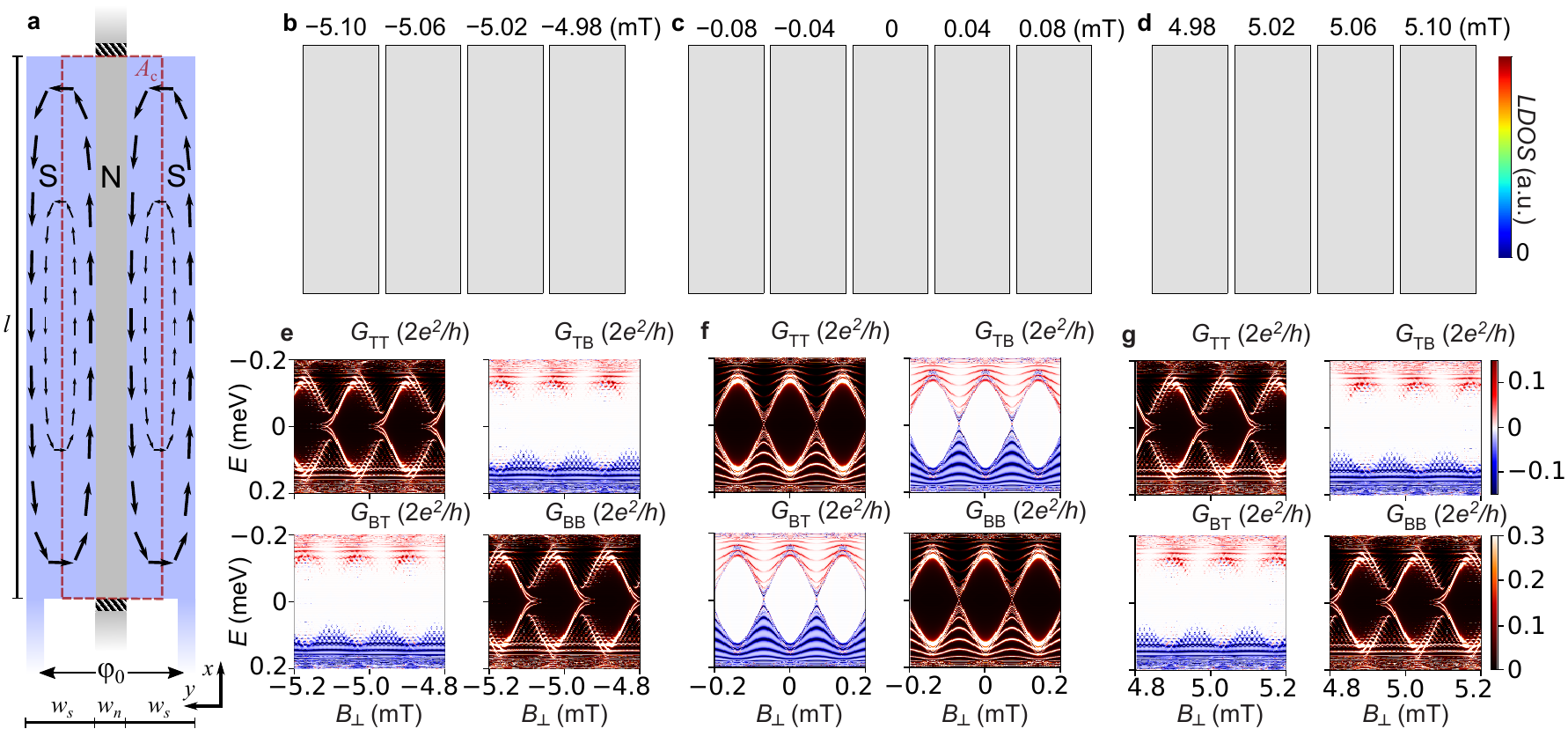}
\caption{\label{fig02} 
{\bf Theory.} (a) Schematic of the device showing supercurrent profile in the superconducting leads, of width $w_s$ and length $l$, at finite $B_\perp$. Arrows correspond to the gauge-invariant local supercurrent density $\vec{j}$. A global phase difference $\phi_0$ is imposed between the S leads by a superconducting loop (not shown). The normal barrier N, of width $w_n$, is connected by tunnel barriers to normal leads at the two ends for conductance calculations. The red dashed lines indicate the central area of the junction that determines the relative local phase difference between the top and bottom ends. (b) -- (d) LDOS of the lowest-energy Andreev bound state around (b) $B_\perp = -5$~mT, (c) $B_\perp=0$ and (d) $B_\perp = +5$~mT, in steps of $\Delta B_\perp = 0.04$~mT.
(e)--(g) Calculated conductance matrix at finite out-of-plane magnetic field around (e) $B_\perp = -5$~mT, (f)$B_\perp =0$, and (g) $B_\perp =5$~mT, respectively. }
\end{figure*}

\begin{figure}[t]
\includegraphics[width=0.5\textwidth]{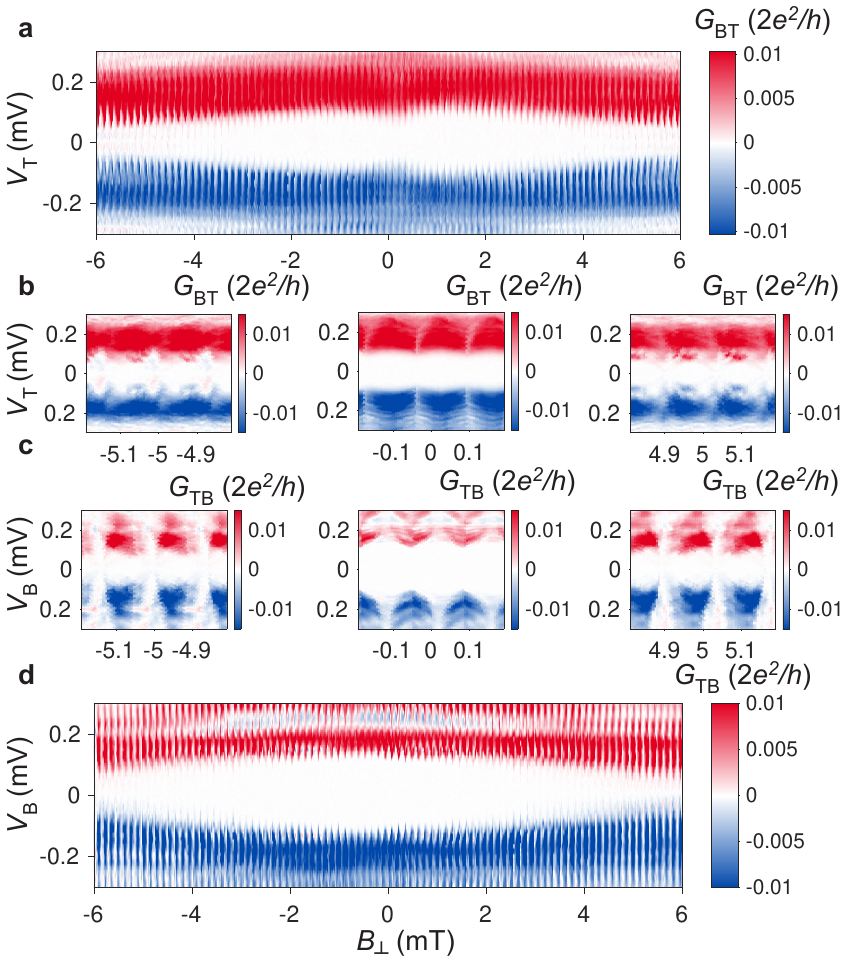}
\caption{\label{fig03} {\bf Nonlocal differential conductance as a function of perpendicular magnetic field.} Nonlocal differential conductances (a) $G_{\rm BT}$ and (d) $G_{\rm TB}$ , as a function of the out-of-plane magnetic field $B_\perp$. The nonlocal gap is periodically modulated with the loop flux. The magnitude of the nonlocal gap at both ends is diminished as $|B_\perp|$ increases. (b) $G_{\rm BT}$ and (c) $G_{\rm TB}$ measured for $\sim$~3 flux lobes centered around (left) $B_\perp=-5$~mT (center) $B_\perp=0$ and (right) $B_\perp=+5$~mT. Compared to the local conductances, the nonlocal conductances are relatively phase-symmetric at all values of $B_\perp$. Also, sub-gap states that approach zero-energy in local conductances (Fig.~\ref{fig01}) are absent in the nonlocal spectrum}.
\end{figure}

This conductance fingerprint is captured by theoretical simulations. The combination of phase-biasing and magnetic field penetrating the Josephson junction creates a phase texture within the two S leads, which we obtain from a  Ginzburg-Landau calculation (see Supplemental Material). As a result, the proximity induced gap in the N region is spatially modulated. Andreev bound states are trapped at positions along the junction where the local phase difference $\sim(2n+1)\pi$, $n$ being an integer. For every flux quantum ($\Phi_0=h/2e$) that enters (exits) the superconducting loop, localized Andreev bound states are pumped from the outer (inner) to the inner (outer) junction end, creating phase-inversion-asymmetric conductance patterns at the two ends. Since these modes are spatially localized, they do not appear in the nonlocal conductance spectra. When the magnetic field through the junction center (defined later) exceeds $\Phi_0$, well-localized Josephson vortices are formed.

Figure~\ref{fig01}(a) shows an electron micrograph of one of the devices, along with a schematic electrical circuit. The device is fabricated on a molecular-beam-epitaxy grown heterostructure with an InAs quantum well separated from a top Al layer by an In$_{0.75}$Ga$_{0.25}$As barrier. A combination of wet etching of the Al layer and deep wet etching of the semiconductor stack is used to define the superconducting loop, the Josephson junction and the mesa with a U-shaped trench. A patch of the mesa (with Al removed) within the loop is contacted by a layer of Ti/Au to form an internal sub-micron ohmic contact to enable bottom-end tunneling spectroscopy. A layer of HfO$_2$, grown by atomic layer deposition (ALD) and patterned in a rectangular shape, is used to isolate the Ti/Au layer from the superconducting loop and the conducting mesa. A second layer of HfO$_2$ is deposited globally followed by the deposition of Ti/Au gates for electrostatic control of the junction and the QPCs. The carrier density in the normal barrier of the JJ (width $w_n=100$~nm, length $l=1.6\,\mu$m) is controlled by energizing gate voltage $V_1$. Gate voltage $V_{\rm sc}$ controls the carrier density in the semiconductor underneath the superconducting leads. Split gates controlled by voltages $V_{\rm tqpc}$ and $V_{\rm bqpc}$ electrostatically define constrictions at the top and bottom of the junction to serve as QPC tunnel barriers. Additional gates controlled by $V_{\rm top}$ and $V_{\rm bot}$ control densities in the normal regions outside the QPCs, and are typically fixed at $\sim 100$~mV. Here we focus on results from Device 1; qualitatively similar results are obtained in Device 2 (see supplementary material Figs.~\ref{suppfigS8},~\ref{suppfigS9}, and  ~\ref{suppfigS10}).

We first investigate local tunneling spectra at the two ends as a function of perpendicular magnetic field as shown in Fig.~\ref{fig01}(b) (top end) and (e) (bottom end). We observe a superconducting gap at both ends that is periodically modulated with a periodicity of $B_\perp \simeq 0.14$~mT, corresponding to $\Phi_0$ through the superconducting loop. The spectra also show a large scale structure with respect to $B_\perp$, where the amplitude of gap modulation is suppressed with $|B_\perp|$ (see also Figs.~\ref{suppfigS6} and \ref{suppfigS7}). 

Focusing on the large scale gap structure, we observed that individual flux lobes acquire significant phase asymmetry as $|B_\perp|$ was increased. This is represented in Figs.~\ref{fig01}(c) and (d). For flux lobes centered around $B_\perp=-5$~mT, the top-end local differential conductance $G_{\rm TT}$ shows a set of arrowhead-shaped features that approach zero-bias at the left (more negative field) end of the lobe, and become maximally gapped at the right (less negative field) end. This pattern is reversed for the bottom-end conductance spectra, $G_{\rm BB}$. Investigating the structure of the flux lobes at $B_\perp=+5$~mT shows that the sense of this asymmetry-in-flux is reversed at both ends. Flux lobes centered around $B_\perp=$~0 are left-right symmetric, that is, the gap is modulated symmetrically within each flux lobe at both ends. 

To help understand this behavior, we perform theoretical simulations of the device. When subjected to a perpendicular magnetic field, identical Meissner supercurrents are set up in the superconducting leads, leading to a spatially varying phase profile (see supplementary material for details). We use Ginzburg-Landau equations to calculate the phase profile within the gauge $\vec{A} = - y B_\perp \hat{x}$. The phase configurations in the left ($\phi_{L}$) and right leads ($\phi_{\rm R}$) obtain a $B_\perp$ dependent gradient given by $d\phi_{\rm L/R}(x)/dx = \pm \pi d_{\rm S} B_\perp/ \Phi_0$, where $d_{\rm S}=(w_n+w_s) = 300$~nm is the center-to-center distance between the two superconducting leads. The effect of the magnetic field through the superconducting loop is modeled as a global phase difference, $\phi_0=2\pi A_{\rm L} B_\perp /\Phi_0$ between the two leads, where $A_{\rm L} \simeq 15~\mu \rm m^2$ is the effective loop area. As a result, the gauge-invariant phase difference across the junction is given as $\Delta\phi(x)=\phi_0+ (2\pi d_{\rm S} B_\perp/\Phi_0)x$.
The central area of the junction $A_{\rm c} = d_{\rm S} l$ determines the total phase winding along the junction $\Delta\phi(l) - \Delta\phi(0) = (2\pi B_\perp A_{\rm c}/\Phi_0)$. Each $2\pi$ winding of the local phase difference corresponds to a Josephson vortex in the junction \cite{cuevas2007magnetic,roditchev2015direct}. 

In the model, ABSs are trapped in nodes of the local phase difference, when $\Delta \phi(x_0) = \pi$ at a position $x_0$ along the junction. The node position, $x_0$, can be controlled by the global phase bias, $\phi_0$, while the localization length of the ABS along the junction, $\xi_B= \sqrt{\Phi_0/B_{\perp}} \sqrt{\xi_P/d_S}$, where $\xi_P$ is the proximity induced coherence length (see supplementary material). At large $B_\perp$, $\xi_B \leq l$, and well developed Josephson vortices are formed at the nodes. Using exact diagonalization of a minimal tight-binding model of the Josephson junction, we find that the nodes bind low-energy Andreev bound states (ABS). In Figs.~\ref{fig02}(a, b), we show the local-density-of-states (LDOS) map of the lowest energy ABS trapped within the junction, formed at $B_\perp=-5.02$~mT. Variation of the loop flux by $B_\perp$ changes the position of the node, as represented by the LDOS maps in Fig.~\ref{fig02}(b) at different values of magnetic field centered around $B_\perp=-5$~mT. At $|B_\perp| \simeq 0$, ABSs are delocalized over the entire length of the junction [Fig.~\ref{fig02}(c)], and exhibit weak spatial modulations. Unlike Fraunhoffer oscillations, where the critical Josephson current is minimal when the junction traps a flux quantum, we do not observe distinctive spectral signatures associated with this effect (see Figs.~\ref{suppfigS2},~\ref{suppfigS4},~\ref{suppfigS6}, and~\ref{suppfigS8}).

These ABSs produce distinctive conductance signatures. We attach normal leads to the two ends of the junction, and evaluate the 2$\times$2 conductance matrix as a function of $B_\perp$ [Figs.~\ref{fig02}(e)-(g)]. Around $B_\perp=-5$~mT, Figs.~\ref{fig02}(e), local conductances, $G_{\rm TT}$ and $G_{\rm BB}$, display phase-inversion-asymmetric lobes with opposite asymmetries at the two ends. A gap minimum in $G_{\rm TT}$ and $G_{\rm BB}$, is observed when the local phase difference $\Delta \phi(x) \sim \pi$ at the top ($x=l$) and bottom ($x=0$) ends respectively. This happens at different magnetic field values, with $\Delta B_\perp=A_{\rm c} B_{\perp} /A_{\rm L}$.

Around each gap minimum, the lowest energy states fade at the top and reappear at the bottom as the magnitude of the magnetic field is increased. These are the strongly localized, lowest-energy modes trapped in the phase-node. In contrast, nonlocal conductances, $G_{\rm TB}$ and $G_{\rm BT}$, remain comparatively phase-inversion-symmetric and do not have pronounced gap minima. The main contribution to nonlocal conductance comes from states that are extended throughout the junction~\cite{DanonNonlocal,MenardNonlocal}, and therefore largely unaffected by local phase differences. At $B_\perp = 0$ [Fig.~\ref{fig02}(f)], all the junction modes are extended and the flux lobes are symmetric in both local and nonlocal conductances. 

Even though our conductance calculations are performed in the clean limit, in the diffusive limit we expect similar results with a renormalized superconducting coherence length in the Al/InAs heterostructure \cite{roditchev2015direct}. Furthermore, we expect that spin-orbit coupling is not relevant for the central features of our data (see supplementary material for exact diagonalization of the vortex ABS, which are almost spin degenerate).

We next consider experimental measurement of nonlocal differential conductance (see supplementary material and Ref.~\cite{BanerjeeNonlocal} for details). As seen in  Fig.~\ref{fig03}, the phase-inversion-asymmetry that was present in  local conductances is strongly suppressed in the nonlocal spectra. For example, flux lobes centered around $B_\perp= \pm 5$~mT, shown in Figs.~\ref{fig03}(b, c), display weak asymmetry within each flux lobe. Furthermore, states that close the spectral gap in the local spectra [Figs.~\ref{fig01}(c, d)], are absent in nonlocal spectra. 

Reasonable agreement between theoretical and experimental conductance matrix signatures supports our interpretation of spatial manipulation of ABSs with superconducting phase texture. In the presence of a finite in-plane magnetic field, Josephson junctions with strong spin-orbit coupling, such as ours, may host spin-split Andreev bound states and topologically protected Majorana zero modes~\cite{HellFlensberg,Pientka}. Spatial manipulation of these states may be realized using the same scheme, enabling non-abelian braiding and fusion-rule experiments~\cite{Stern2019Fractional,ShabaniXjunction,Hegde2020topological}. Future work may focus on this direction. 

We thank Geoff Gardner and Sergei Gronin for contributions to materials growth, and  Asbj{\o}rn Drachmann for assistance with fabrication. We acknowledge a research grant (Project 43951) from VILLUM FONDEN.
MG acknowledges support by
the European Research Council (ERC) under the European Union’s Horizon 2020 research and innovation program under grant agreement No.~856526, and from the
Deutsche Forschungsgemeinschaft (DFG) project grant
277101999 within the CRC network TR 183 (subproject
C01), and from the Danish National Research Foundation, the Danish Council for Independent Research $\vert$ Natural Sciences.

\bibliography{Vortex-planarJJ.bib}

\clearpage

\onecolumngrid
\appendix
\begin{center}
{\bf Supplementary Material for \\Control of Andreev bound states using superconducting phase texture}
\end{center}
%\makeatletter
%\renewcommand{\fnum@figure}{\figurename~S\thefigure}
%\makeatother
%\setcounter{figure}{0}   

\setcounter{equation}{0}
\renewcommand{\theequation}{S\arabic{equation}}
\setcounter{figure}{0}
\renewcommand{\thefigure}{S\arabic{figure}}
\setcounter{section}{0}
\renewcommand{\thesection}{S\Roman{section}}

\vspace{10 pt}

{\bf Wafer structure:} The wafer structure used in this work consists of an InAs two-dimensional quantum well in epitaxial contact with Aluminum, grown by molecular beam epitaxy. The wafer was grown on an insulating InP substrate and comprises a 100-nm-thick In$_{0.52}$Al$_{0.48}$As matched buffer, a 1$\mu$m thick step-graded buffer realized with alloy steps from In$_{0.52}$Al$_{0.48}$As to In$_{0.89}$Al$_{0.11}$As (20 steps, 50 nm/step), a 58 nm In$_{0.82}$Al$_{0.18}$As layer, a 4 nm In$_{0.75}$Ga$_{0.25}$As bottom barrier, a 7 nm InAs quantum well, a 10 nm In$_{0.75}$Ga$_{0.25}$As top barrier, two monolayers of GaAs and a 7 nm film of epitaxially grown Al. The top Al layer was grown without breaking the vacuum, in the same growth chamber.  Hall effect measurements performed in Hall bar devices of the same material, with Al etched away, indicated a peak electron mobility peak $\mu= 43,000$~ cm$^2$/Vs at a carrier density of $n=8 \times 10^{11}$~cm$^{-2}$, corresponding to a peak electron mean free path of  $l_e\sim$~600 nm suggesting that our devices are quasi-ballistic along the length $l \simeq 3 l_e$ and ballistic in the width direction $w_n \simeq l_e/6$. Transport characterization of an etched Aluminum Hall bar revealed an upper critical field of 2.5~T, indicating that the Al layer loses superconductivity at a field much larger than the collapse of the induced superconducting gap at $\sim$~0.5~T.   

\vspace{10 pt}

{\bf Device fabrication:} Devices were fabricated with standard electron beam lithography techniques. Devices on the same chip were electrically isolated from each other using a two-step mesa etch process, first by removing Al with Al etchant Transene D, and then a standard III-V chemical wet etch using a solution comprising H$_2$O : C$_6$H$_8$O$_7$ : H$_3$PO$_4$ : H$_2$O$_2$ (220:55:3:3) to etch the mesa until a depth of $\sim$300~nm. This step also defined the U-shaped trench and the patch of mesa that eventually formed the submicron ohmic contact. In the next lithography step, the Aluminum layer was selectively removed leaving behind the Josephson junction with a flux loop. Particularly, this step also removed Al from the internal ohmic mesa patch. A patterned layer of dielectric, comprising 15 nm thick HfO$_2$ grown at 90$^\circ$C using atomic layer deposition (ALD), was then deposited to galvanically isolate the ohmic Ti/Au layer from the rest of the device. This was followed by the deposition of Ti/Au layers (5~nm/300~nm) for the inner ohmic contact. Next, a global layer of 15 nm thick HfO$_2$ was deposited over the entire sample to serve as the gate dielectric. Gates were defined using electron beam lithography followed by e-beam evaporation of Ti/Au layers of thickness (5nm/20nm) for finer structures and (5nm/350nm) for the bonding pads.

\vspace{10 pt}

{\bf Three-terminal electrical transport measurements:} Electrical transport measurements were performed in an Oxford Triton dilution refrigerator at a base temperature of $\sim$20 mK using conventional low frequency AC lock-in techniques. The superconducting loop was grounded by connecting it to the fridge ground through a low resistance path ($\sim$0.9~k$\Omega$). The top and the bottom ohmic contacts were connected to low-impedance current-to-voltage converters, through low resistance paths ($\sim$1.5--2.0~k$\Omega$). The grounding pin of each current-to-voltage converter was biased through a combination of AC+DC voltages $V_{\rm t(b)}+V_{\rm T(B)}$, at the top and bottom ends respectively. The AC biases, $V_{\rm t}$ and $V_{\rm b}$, were generated by two lock-in amplifiers, with the same excitation amplitude (3~$\mu$V), but different frequencies, $f_{\rm t}=31.5$~Hz and $f_{\rm b}=77.5$~Hz respectively. The DC biases $V_{\rm T(B)}$ were generated using two low-noise DC voltage sources. The voltage output of each current-to-voltage converter was measured using two lock-in amplifiers, operating at frequencies $f_{\rm t}$ and $f_{\rm b}$, requiring a total of four  lock-in amplifiers for each element of the 2$\times$2 conductance matrix.

\vspace{10 pt}

{\bf Phase-biasing and magnetic field alignment:}  Magnetic field to the sample is applied using a three-axis ($B_x$,$B_y$,$B_z$)=(1T,1T,6T) vector rotate magnet. The sample is oriented with respect to the vector magnet such that $B_x$ is nominally parallel to $B_\perp$ [Fig.~\ref{fig01}(a)]. 

\vspace{10 pt}

\begin{figure}[t]
    \centering
    \includegraphics{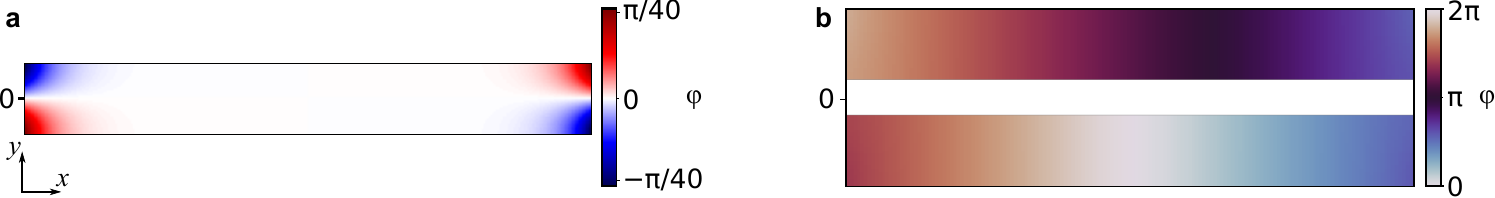}
    \caption{\textbf{Theory: Superconducting phase profile.} Calculated phase profile $\phi(x,y)$ from the Ginzburg-Landau equation along Ref. \cite{clem2010} in the gauge $\vec{A} = - y B_\perp \hat{x}$, (a) for a single superconducting lead centered at $y = 0$ at $B_\perp = -5.02$~mT and (b) for the planar Josephson junction consisting of two leads centered at $y = \pm d_{\rm S}/2$ at $B_\perp = -5.02$~mT including the phase bias $\phi_0 = 2\pi \frac{B_\perp A_{\rm L}}{\Phi_0}$ due to the loop. The additional phase gradients in the two leads arise due to the displacement of the superconductors away from $y = 0$ in the gauge $\vec{A} = - y B_\perp \hat{x}$. This phase profile corresponds to the supercurrent profile sketched in Fig.~\ref{fig02}~(a) in the main text via relation \eqref{eq:supercurrent_density}.}
    \label{suppfigS1}
\end{figure}

{\bf Phase profile in the superconducting leads:}
To determine the phase profile in the superconducting leads forming the wide Josephson junction, we employ the solution of the Ginzburg-Landau equations for thin films considered in Ref.~\onlinecite{clem2010}. This solution requires the following assumptions:
(i) the superconducting film (thickness $d=7$~nm) is thinner than the effective penetration depth for thin films $\lambda_\text{eff} = \lambda_L \sqrt{\xi_0 / d} \simeq 240$~nm \cite{tinkham_introduction_2004}, where $\lambda_L=16$~nm the London penetration depth, $\xi_0 = 1600$~nm BCS coherence length of Al \cite{kittel2004}, (ii) the Josephson current density is small compared to the Meissner current density such that changes in the supercurrent and phase profile due to coupling to the two-dimensional electron gas can be neglected, (iii) the magnetic self-field generated by the Meissner currents is much smaller than the applied field, (iv) the current current density in the superconducting leads is much smaller than the critical current density such that the reduction of the magnitude of the order parameter can be neglected. We verify these assumptions from the result, which is displayed in Fig.~\ref{suppfigS1} for the gauge choice $\vec{A} = -y B_\perp \hat{x}$ for $B_\perp = -5.02$~mT. 

 A conservative estimate for the typical supercurrent density flowing in the superconducting leads in response to an applied perpendicular field is given by the supercurrent density flowing along the edge of the junction.
The gauge-invariant supercurrent density is given by \cite{clem2010}
\begin{equation}
    \vec{j} = -\frac{1}{\mu_0 \lambda_\text{eff}^2} \left( \vec{A} + \frac{\Phi_0}{2\pi} \nabla \phi \right).
    \label{eq:supercurrent_density}
\end{equation}
Neglecting the contribution from the superconducting phase gradient $\nabla \phi(x,y)$, which vanishes in the bulk of a rectangular superconducting lead with large aspect ratio for our gauge choice $\vec{A} = -y B_\perp \hat{x}$, 
 {\it c.f.} Fig.~\ref{suppfigS1}(a), the supercurrent density at the edge of the junction is given by
  $j_{y}|_\text{edge}(B_\perp)=-\frac{ w_{\rm S}}{4\mu_{0}\lambda_{\text{eff}}^{2}}B_{\perp}$.
  At $B_{\perp}=5$~mT, the Meissner current density averaged over the thickness $d$ of the Al film $j_{y}|_\text{edge}(5\text{mT}) d=2\times 10^{-8}$ is
  much larger than the critical current density $j_{c}=\frac{e\Delta_{0}}{\hbar}\frac{1}{\lambda_{F}}=3.2\times  10^{-10}$~A/nm of the junction.
  Also, the Meissner supercurrent density $j_{y}|_\text{edge}(5\text{mT})=3\times 10^{-9}$~A/nm$^2$ is much smaller than the pair-breaking critical current density of the Al film $j_{c}^{\text{Al}}\simeq10^{-7}$~A/nm$^2$ \cite{romijn1982}. Thus assumptions (ii) and (iv) are satisfied.
 Approximating the continuous current density in the superconducting leads by a current along an elliptical loop with half axis $a=L/2, b=w_{\rm S}/4$ carrying a current $\bar{I}_{S}=j_{y}w_{\rm S}/4$, the Biot-Savart law yields for the magnetic self-field at the center $\vec{B}_{\text{self}}\simeq\frac{\mu_{0}\bar{I}_{S}}{\pi w_{\rm S}/4}=-\frac{W_{S}}{2\pi\Lambda}B_{\perp}\hat{z}$  with the Pearl length $\Lambda = \frac{2 \lambda_\text{eff}^2}{d} \simeq 16.5\,\mu\text{m}$ and $w_{\rm S}/2\pi\Lambda\simeq0.002$ such that the self-field may safely be neglected.
 
The superconducting leads are displaced by $\pm d_{\rm S}/2 = \pm 150$~nm from $y=0$. In our gauge choice $\vec{A} = - y B_\perp \hat{x}$, the phase-profile in the right / left superconducting lead contains a linear phase gradient term $\phi_{R/L}(x,y) = \phi_{y_c = 0}(x,y) \pm \pi d_{\rm S} x B_\perp / \Phi_0$
to ensure that the gauge-invariant supercurrent profile is identical in the two superconducting leads, where $\phi_{y_c = 0}(x,y)$ is the phase-profile calculated for a rectangular superconducting lead centered at $y = 0$, as shown in Fig.~\ref{suppfigS1}(a). The phase profile including this phase gradient is shown in Fig.~\ref{suppfigS1}(b).
In other words, the phase gradient term $\pm \pi d_{\rm S} x B_\perp / \Phi_0$ ensures that the $j_x$ component of the supercurrent vanishes within the mirror planes of the rectangular superconducting leads, as it should for the supercurrent vortex profile shown in Fig.~\ref{fig02}(a) in the main text. 
In our experimental geometry, the distance $d_{\rm s} = 300$~nm corresponds to the distance of the centers of the Al meander arm closest to the junction, taken at minimal width. 

\vspace{10 pt}

{\bf Theory details:}  
We employ a tight-binding discretization of the spin-orbit coupled two-dimensional electron gas (2DEG) in the InAs quantum well including the coupling to the proximitized Al superconductors in terms of a frequency-dependent self energy $\Sigma(\omega)$, and metallic leads to compute the conductance matrix using the software package \textit{kwant}  \cite{groth_kwant_2014}. 
The discretized tight-binding Hamiltonian is
\begin{align}
\label{eq:S_H_N}
    H & =\sum_{\sigma,\sigma^{\prime}=\uparrow,\downarrow}\sum_{\vec{r}}\left(\left(\frac{\hbar^{2}}{m^{*}a^{2}}-\mu\right)\delta_{\sigma,\sigma^{\prime}}+\sum_{j=x,y,z}\frac{1}{2}g\mu_{B}B_{j}\sigma_{\sigma,\sigma^{\prime}}^{j}\right)\hat{c}_{\vec{r},\sigma}^{\dagger}\hat{c}_{\vec{r},\sigma^{\prime}}\\
	\ & +\sum_{\sigma,\sigma^{\prime}=\uparrow,\downarrow}\sum_{\vec{r}}\left(\left(\frac{\hbar^{2}}{4m^{*}a^{2}}\delta_{\sigma,\sigma^{\prime}}+\frac{i\alpha}{2}\sigma_{\sigma,\sigma^{\prime}}^{y}\right)\hat{c}_{\vec{r}+\vec{a}_{x},\sigma}^{\dagger}\hat{c}_{\vec{r},\sigma}+\left(\frac{\hbar^{2}}{4m^{*}a^{2}}\delta_{\sigma,\sigma^{\prime}}-\frac{i\alpha}{2}\sigma_{\sigma,\sigma^{\prime}}^{x}\right)\hat{c}_{\vec{r}+\vec{a}_{y},\sigma}^{\dagger}\hat{c}_{\vec{r},\sigma}+\text{h.c.}\right) \nonumber
\end{align}
with the effective mass $m^* = 0.023\,m_e$, lattice spacing $a=20$~nm, chemical potential $\mu = 1$~meV, $g$-factor $g=5$, Bohr magneton $\mu_B$, Rashba spin-orbit coupling strength $\alpha = 150\,\text{meV}\AA$, chemical potential $\mu = 1$~meV, Kronecker-delta $\delta_{i,j}$ and Pauli matrices $\sigma^{j}, \ j=x,y,z$ in spin space. 
The orbital effect of the magnetic field is included via the substitution of the hopping terms
$$
\hat{c}_{\vec{r}+\vec{a}_{x/y},\sigma}^{\dagger}\hat{c}_{\vec{r},\sigma}\to e^{-i\frac{e}{\hbar}\int_{\vec{r}}^{\vec{r}+\vec{a}_{x/y}}\vec{A}d\vec{r}}\hat{c}_{\vec{r}+\vec{a}_{x/y},\sigma}^{\dagger}\hat{c}_{\vec{r},\sigma}.
$$
The proximity-induced pairing is included via the self-energy term \cite{McMillanPR1968, HellFlensberg}
\begin{equation}
\label{eq:S_H_S}
    H_\Delta(\omega) = \sum_{\sigma,\sigma^{\prime}=\uparrow,\downarrow}\sum_{\vec{r}}\frac{\Gamma(\vec{r})}{\sqrt{|\Delta_{0}(\vec{r})|^{2}-\omega^{2}}}\left( \Delta_{0}(\vec{r})i\sigma_{\sigma, \sigma^\prime}^y \hat{c}_{\vec{r},\sigma}^{\dagger}\hat{c}_{\vec{r},\sigma^{\prime}}^\dagger +\text{h.c.} -\omega\delta_{\sigma, \sigma^\prime} \right).
\end{equation}
where $\Gamma(\vec{r})$ is a phenomenological parameter describing the hopping rate between 2DEG and superconducting leads and $\Delta(\vec{r}) = |\Delta_0| e^{i \phi(r)}$ is the order parameter in the superconducting lead with phase modulation $\phi(\vec{r})$ calculated from the Ginzburg-Landau equation, and shown in Fig.~\ref{suppfigS1}. The last term proportional to the frequency $\omega$ corresponds to a renormalization of the normal-state Hamiltonian \cite{McMillanPR1968}. 
The system is simulated in a rectangular region of length $l=1600$nm and total width $w = 500$~nm, where the proximity induced pairing is present only within the range $w_n/2 < |y| < w_n/2 + w_s$ with the width of the normal region $w_n = 100$~nm and the width of the superconducting leads $w_s = 200$~nm.
We approximate the magnitude of the order parameter of the parent Al superconductor $|\Delta_0| = 0.2$~meV and the hopping rate $\Gamma = 0.9$~meV as constant within the superconducting leads. The hopping rate is fitted to find good agreement with the experimental data. A large hopping rate $\Gamma > |\Delta_0|$ is required to reduce the localization length of the ABSs such that a clear gap in local conductance around $B_\perp = -5$~mT is observed.

The spectrum and LDOS of the Andreev bound states in the junction are computed by solving numerically the non-linear eigenvalue problem $H_\text{BdG}(\omega) \Psi = \hbar \omega \Psi$ for the frequency dependent Bogoliubov-de Gennes Hamiltonian $H_\text{BdG}(\omega)$ constructed from Eqs. \eqref{eq:S_H_N} and \eqref{eq:S_H_S} \cite{IstasSci2018}. To compute the conductance matrix, we attach normal leads at the top and bottom of the normal part of the junction. The conductance from the leads into the simulation region is set to $G = 0.3 \frac{e^2}{h}$ by reducing the hopping terms between leads and simulation region. This approximates a quantum point contact in the tunneling regime as present in the experiment.

Figures \ref{suppfigS3} and \ref{suppfigS5} display the calculated conductance matrix for further windows of the perpendicular magnetic field between $B_\perp = -2.15$~mT and $B_\perp = -12.9$~mT. Figure \ref{suppfigS4} displays the calculated conductance matrix over a large perpendicular magnetic field range from $B_\perp = 0$ to $10$~mT.

\vspace{10 pt}

\begin{figure}
    \centering
    \includegraphics[width=\columnwidth]{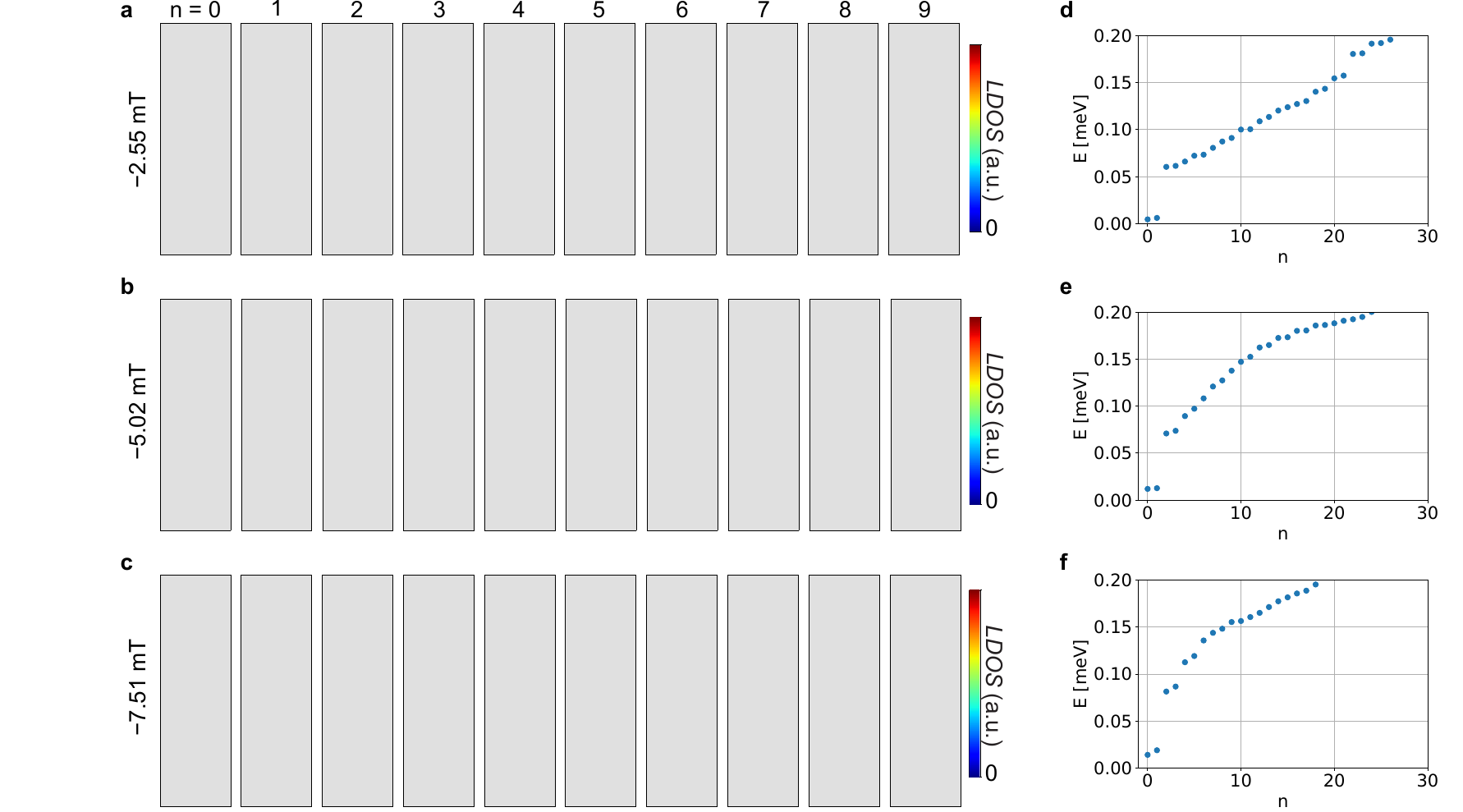}
    \caption{\textbf{Theory: Local Density of States.} (a)-(c) Local density of states (LDOS) of the ten lowest-energy Andreev bound states at $B_\perp = -2.55$~mT, $-5.02$~mT, and $-7.51$~mT, respectively. (d)-(f) Corresponding eigenenergies of the Andreev bound states in the junction at the respective perpendicular field. }
    \label{suppfigS2}
\end{figure}

\begin{figure}[h]
    \centering
    \includegraphics[width=1\textwidth]{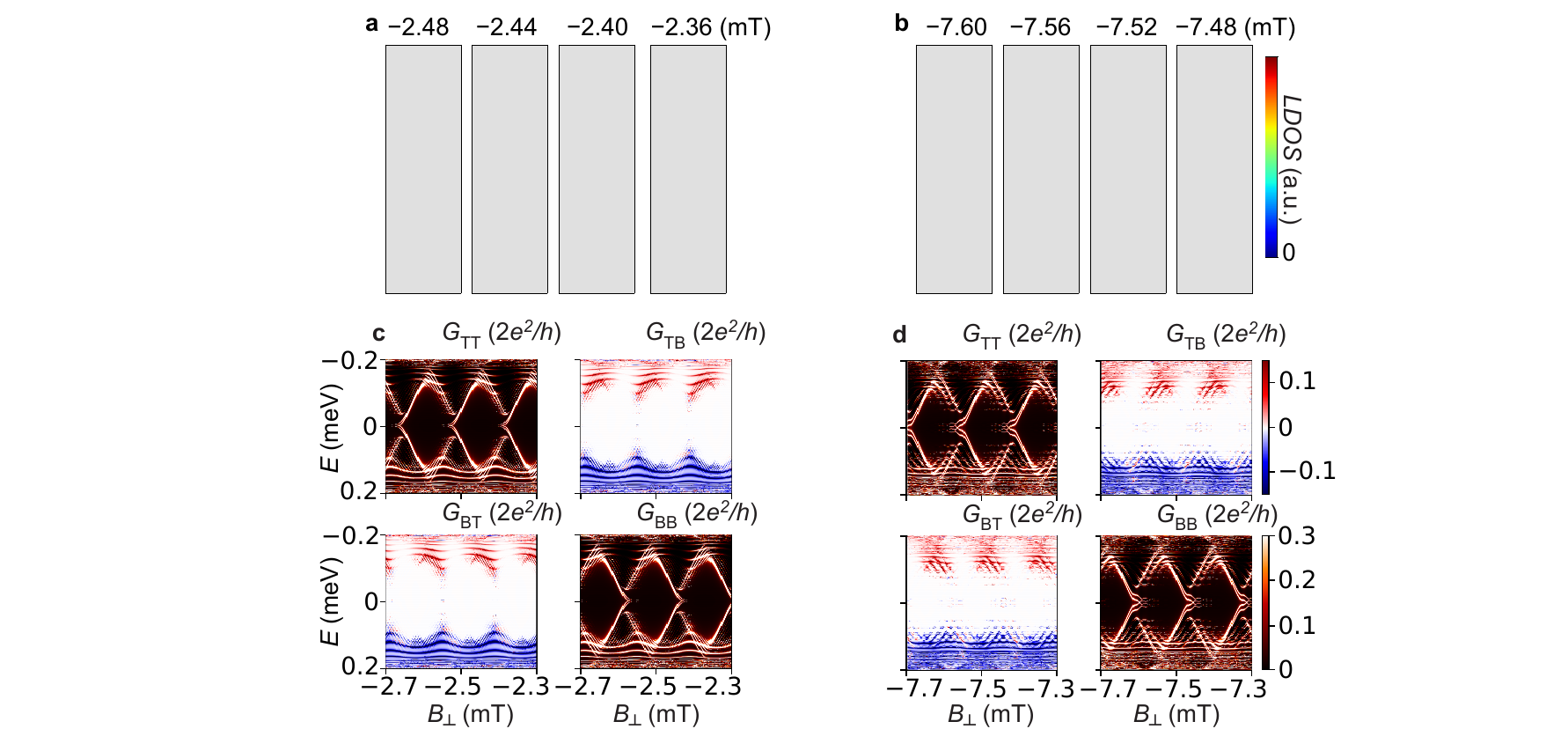}
    \caption{\textbf{Theory: Density of States and Spectra.} (a), (b) LDOS of the lowest-energy Andreev bound state around $B_\perp = -2.5$~mT and $-7.5$~mT in steps of $\Delta B_\perp = 0.04$~mT. (c), (d) Calculated conductance matrix at finite out of plane magnetic field around $B_\perp = -2.5$~mT and $-7.5$~mT, respectively.}
    \label{suppfigS3}
\end{figure}

\begin{figure}[h]
    \centering
    \includegraphics[width=1\textwidth]{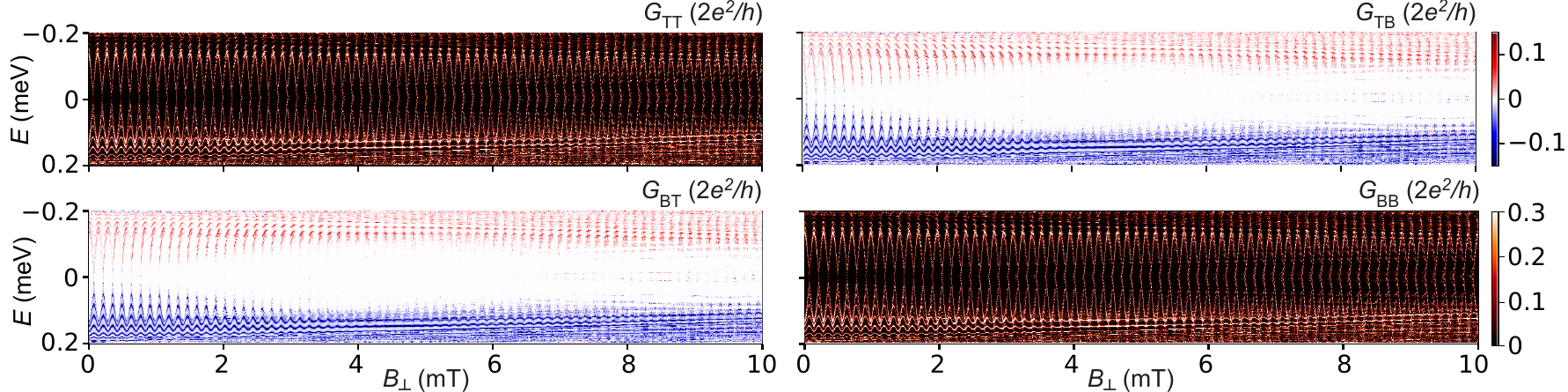}
    \caption{\textbf{Theory: Spectra Over Extended Field Range.} Calculated conductance matrix at finite out of plane magnetic field.}
    \label{suppfigS4}
\end{figure}

\begin{figure}[h]
    \centering
    \includegraphics[width=1\textwidth]{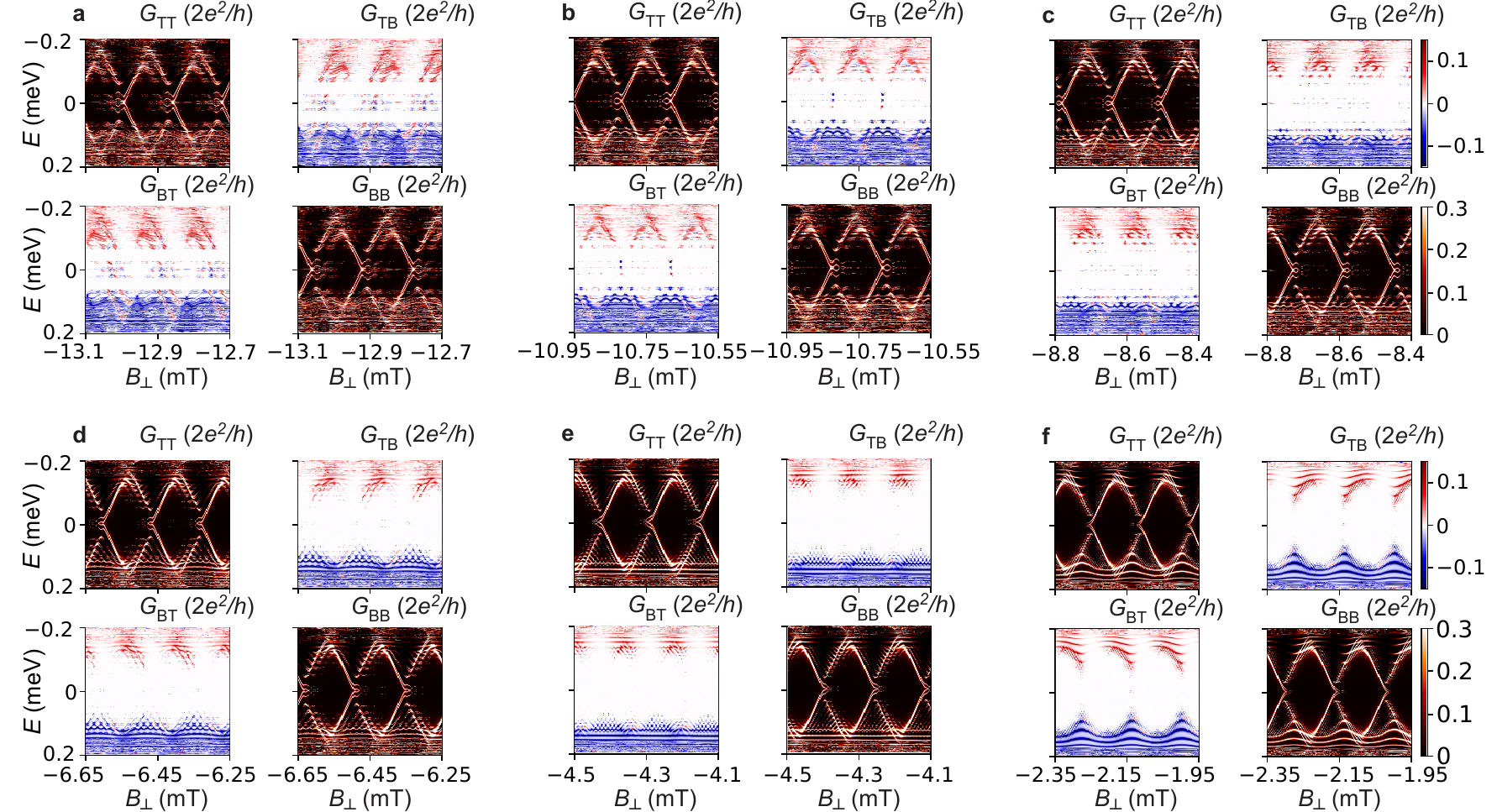}
    \caption{\textbf{Theory: Spectra.} (a) - (f) Calculated conductance matrix at finite out of plane magnetic field around $B_\perp = -12.9$~mT, $B_\perp = -10.75$~mT, $B_\perp = -8.6$~mT, $B_\perp = -6.45$~mT, $B_\perp = -4.3$~mT, and $-2.15$~mT corresponding to a flux through the central area of the junction of $2\pi B_\perp A_{\rm c} / \Phi_0 = 3\pi$, $\frac{5}{2} \pi$, $2\pi$, $\frac{3}{2} \pi$, $\pi$, and $\frac{1}{2} \pi$, respectively.}
    \label{suppfigS5}
\end{figure}

\textbf{Andreev bound states at vortex cores:}
Figure \ref{suppfigS2} displays the calculated LDOS and spectrum of the Andreev bound states at $B_\perp = -2.55$~mT, $-5.02$~mT, and $-7.51$~mT, where the local phase difference $\phi_L(x)-\phi_R(x)$ crosses $\pi$ around the center of the junction. This point coincides with the center of a Josephson vortex in case the junction is wide enough to host a full vortex \cite{roditchev2015direct}. At $B_\perp = -2.55$~mT approximately half a Josephson vortex fits in the junction, {\em i.e.} the local phase difference at the top and bottom ends is $\mp \pi/2$. For $B_\perp = -5.02$~mT ($B_\perp = -7.51$~mT) one full vortex (one and half vortices) fits in the junction. For our geometry, we find that the vortex spectrum contains an almost-degenerate pair of low-energy Andreev bound state that separates from the remaining higher-energy vortex modes. The higher energy vortex modes are approximately equally spaced. Increasing the perpendicular field reduces the vortex size via the gradient of the local phase difference $d(\phi_L - \phi_R)/dx = 2\pi d_{\rm S} B_\perp / \Phi_0$, and thereby increases the energy splitting of the vortex Andreev bound states.

The length scale for the extent of the Andreev bound states below the superconducting leads is given by the proximity coherence length $\xi_{\rm P} = \pi \hbar v_F / \Gamma \simeq 270$~nm. The extent of the ABS around the phase node can be estimated from the average gap experienced by this state \cite{Stern2019Fractional}. For a bound state of extent $\xi_{B}$ around the node, the average gap is $\bar{\Delta} = \Gamma B_\perp d_{\rm S} \xi_B 2 \pi / \Phi_0$. This yields the self-consistency relation for the localization length $\xi_{\rm B} = \hbar v_F / \bar{\Delta} = \sqrt{\Phi_0/B_\perp} \sqrt{\xi_{\rm P}/d_{\rm S}}$. For our parameters, $\sqrt{\xi_{\rm P}/d_{\rm S}} \simeq 1$, such that the localization length is given by the magnetic length $\xi_B \approx \sqrt{\Phi_0/B_\perp}$~\cite{cuevas2007magnetic}.

\clearpage

\clearpage

\begin{figure*}[t]
\includegraphics[width=1\textwidth]{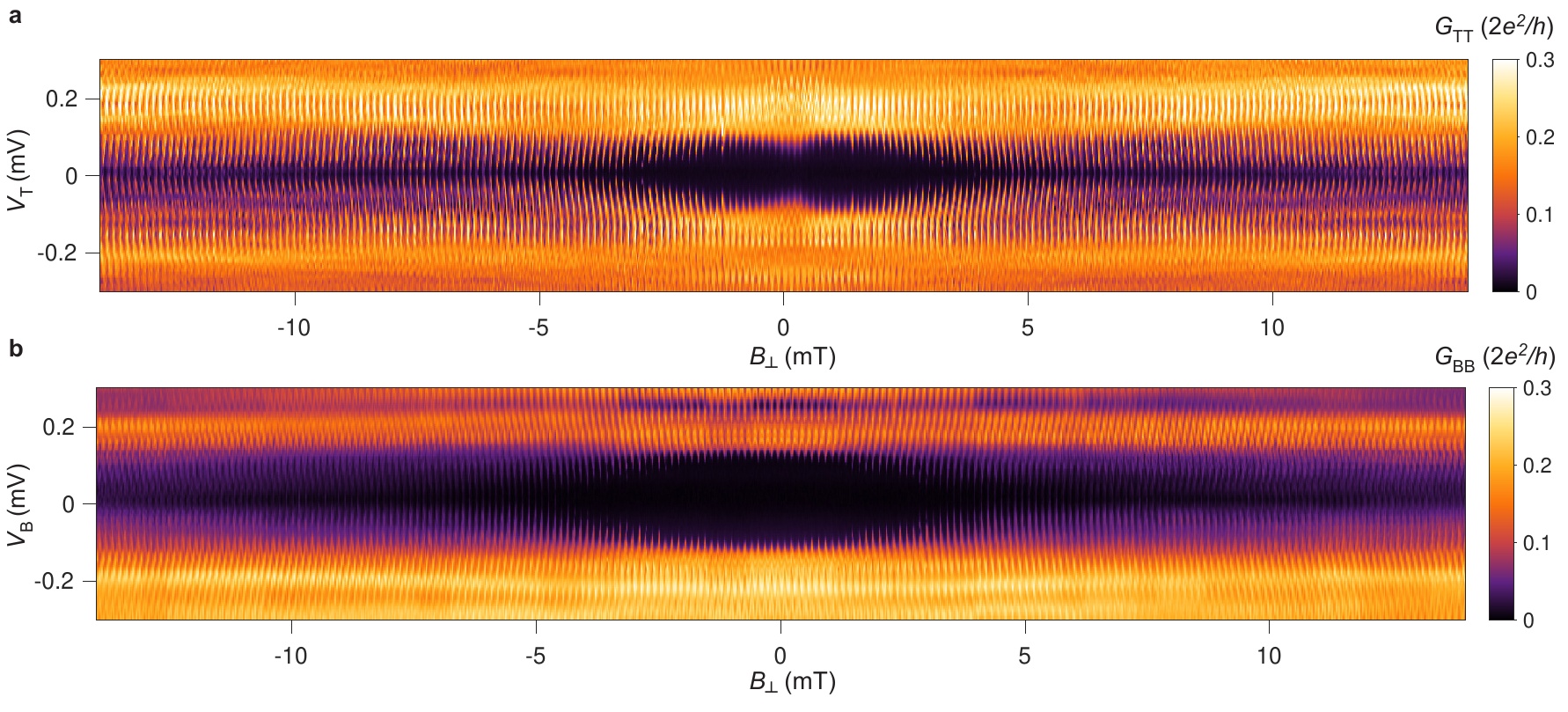} 
\caption{\label{suppfigS6} {\bf Device 1:}~ Local differential conductance measured at the (a) top and (b) bottom ends of the junction, as a function of the out-of-plane magnetic field $B_\perp$. The superconducting gap oscillates periodically, with period comparable with one superconducting flux quantum penetrating the superconducting loop. The magnitude of the gap at both ends is diminished as $|B_\perp|$ increases.}
\end{figure*}

\begin{figure*}[h]
\includegraphics[width=1\textwidth]{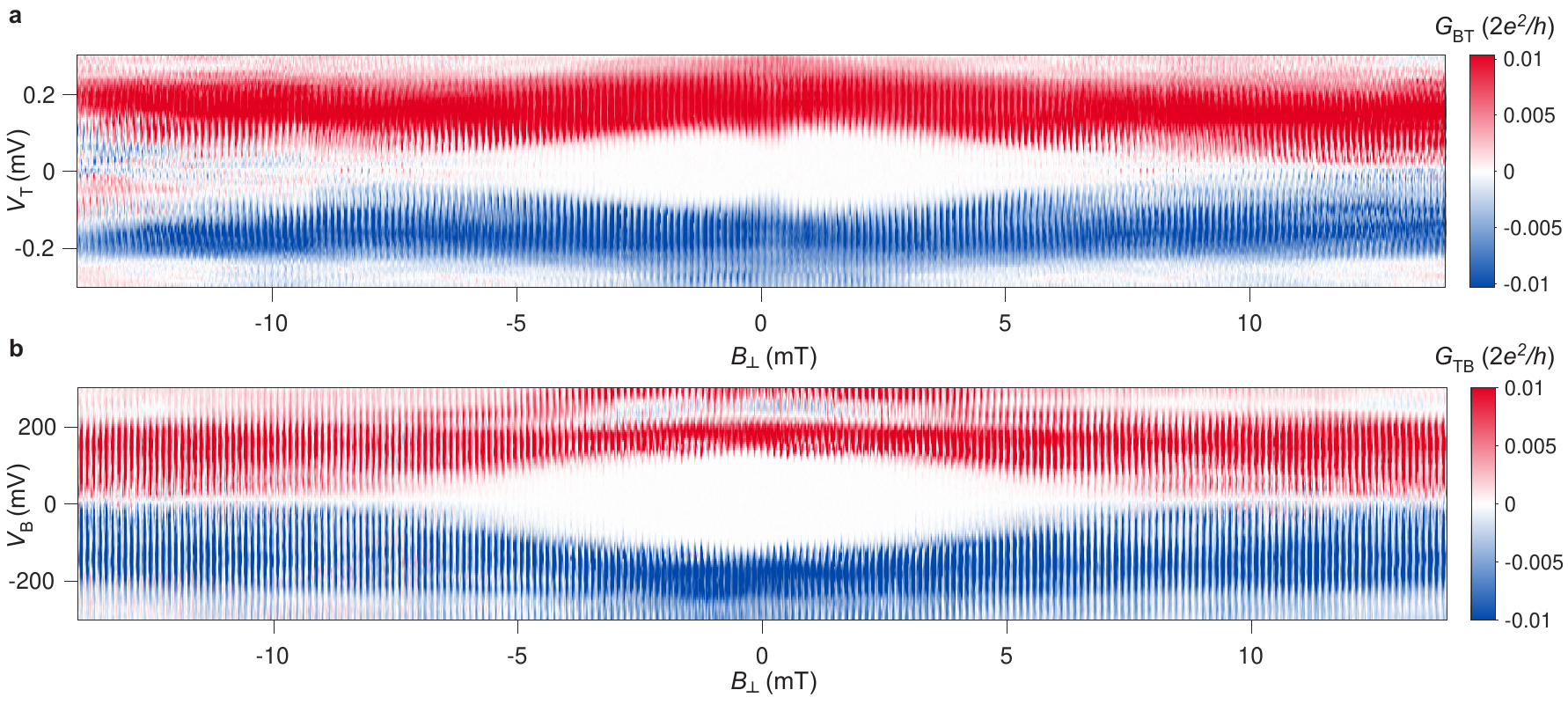} 
\caption{\label{suppfigS7} {\bf Device 1:}~ Nonlocal differential conductances (a) $G_{\rm BT}$ and (d) $G_{\rm TB}$ , as a function of the out-of-plane magnetic field $B_\perp$. The nonlocal gap is periodically modulated with the loop flux. The magnitude of the nonlocal gap at both ends is diminished as $|B_\perp|$ increases.}
\end{figure*}

\begin{figure*}[t]
\includegraphics[width=1\textwidth]{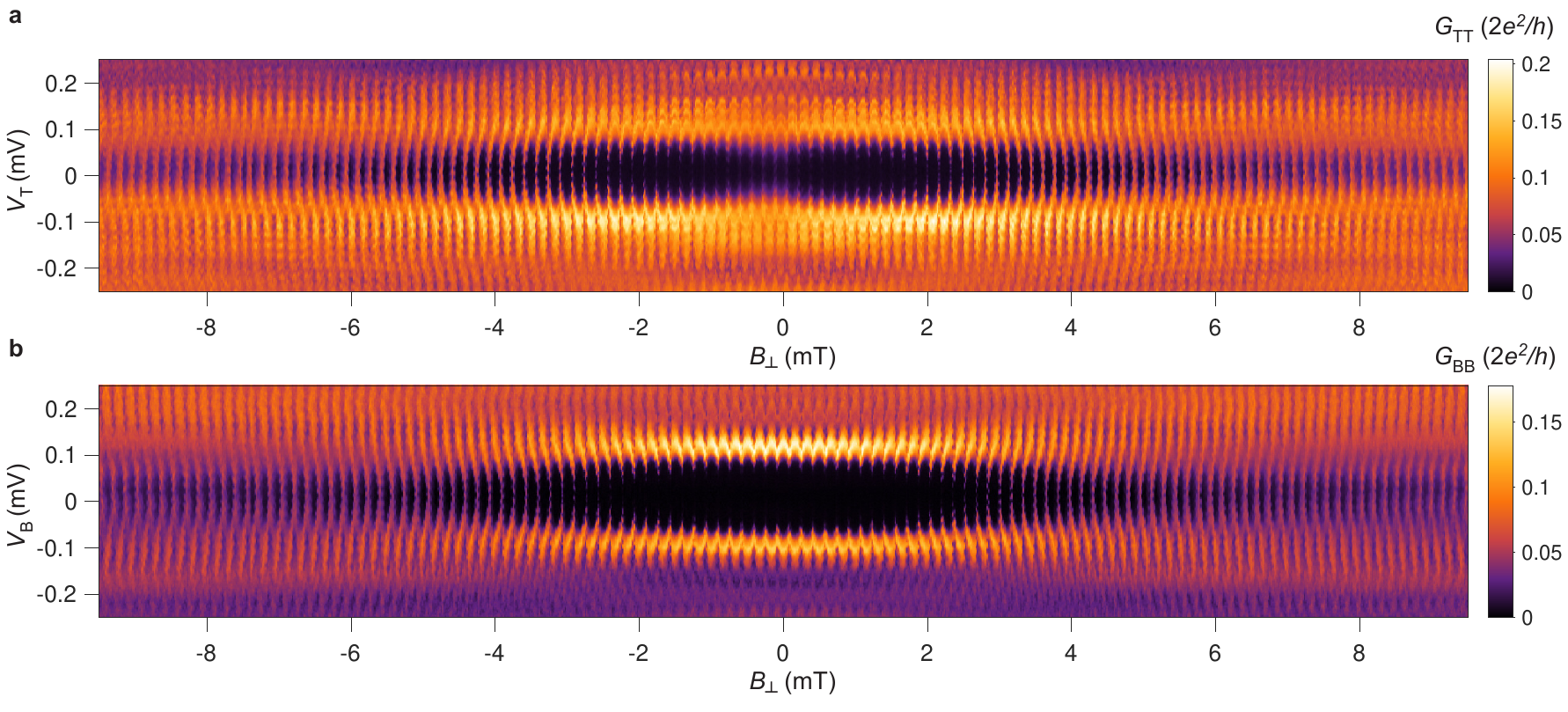} 
\caption{\label{suppfigS8} {\bf Device 2:}~ Local differential conductance measured at the (a) top and (b) bottom ends of the junction, as a function of the out-of-plane magnetic field $B_\perp$. The superconducting gap oscillates periodically, with period comparable with one superconducting flux quantum penetrating the superconducting loop. The magnitude of the gap at both ends is diminished as $|B_\perp|$ increases.}
\end{figure*}

\begin{figure*}[h]
\includegraphics[width=1\textwidth]{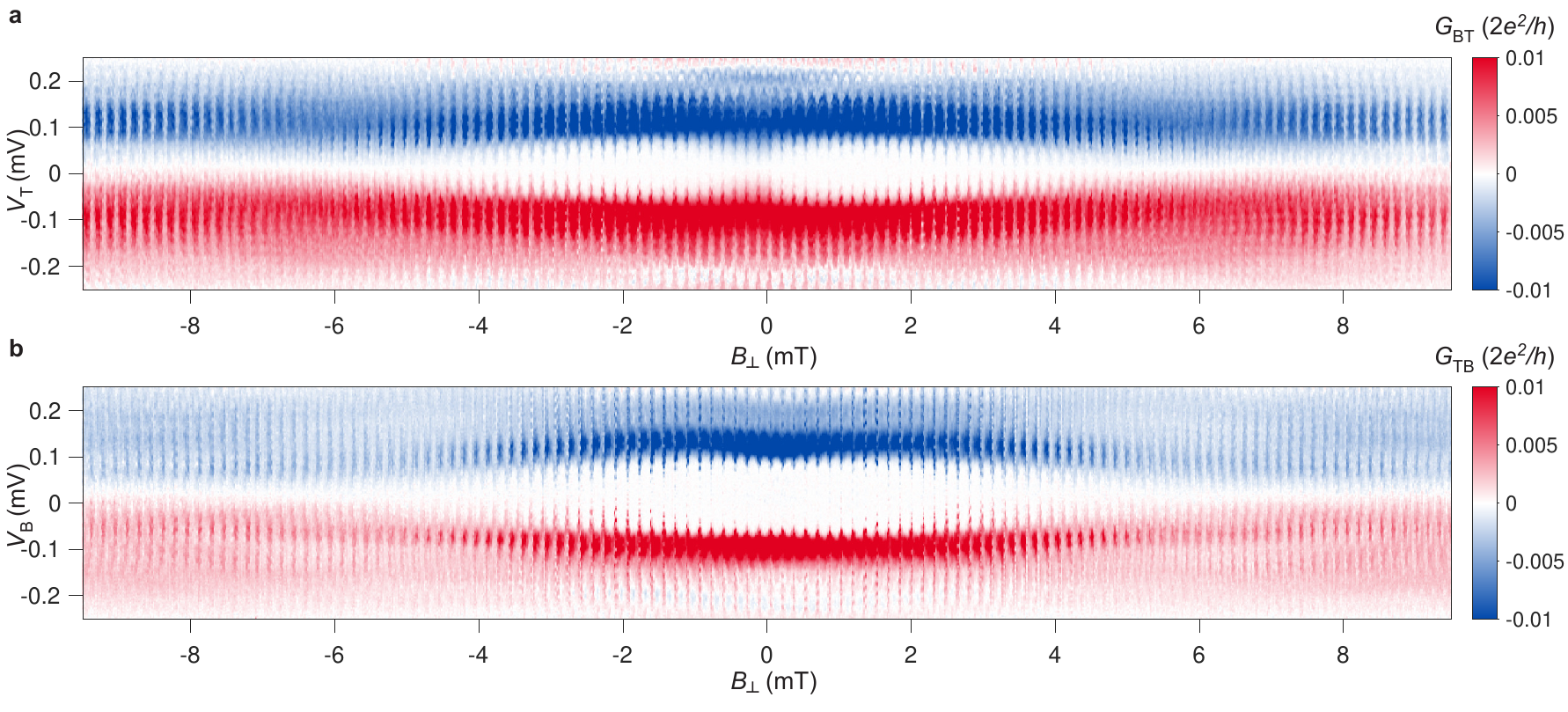} 
\caption{\label{suppfigS9} {\bf Device 2:}~ Nonlocal differential conductances (a) $G_{\rm BT}$ and (d) $G_{\rm TB}$ , as a function of the out-of-plane magnetic field $B_\perp$. The nonlocal gap is periodically modulated with the loop flux. The magnitude of the nonlocal gap at both ends is diminished as $|B_\perp|$ increases.}
\end{figure*}

\begin{figure*}[ht]
\includegraphics[width=1\textwidth]{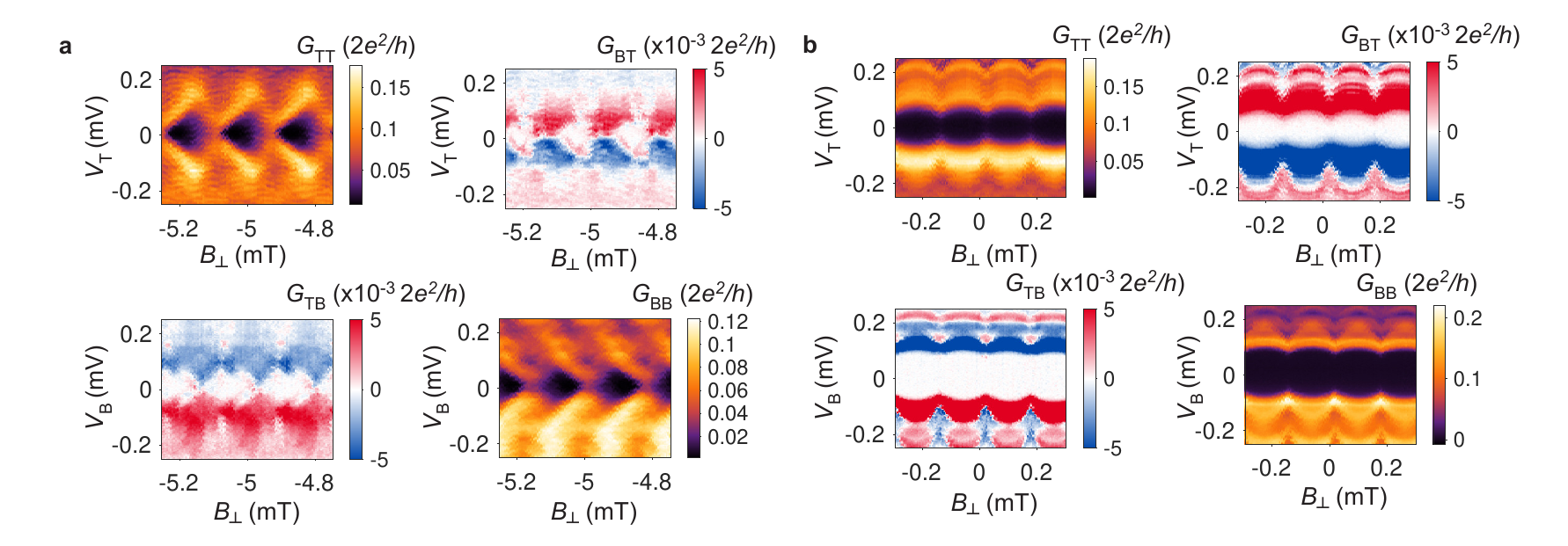} 
\caption{\label{suppfigS10} {\bf Device 2:}~ Conductance matrix showing $\sim$~3 flux lobes centered around (a) $B_\perp=-5$~mT and (b) $B_\perp=0$.}
\end{figure*}

\end{document}